\documentstyle[epsfig,11pt]{article}

\def\roughly#1{\mathrel{\raise.3ex\hbox{$#1$\kern-.75em%
\lower1ex\hbox{$\sim$}}}}

\def\lsim{\roughly<}
\def\gsim{\roughly>}
\def\be{\begin{eqnarray}}
\def\ee{\end{eqnarray}}

\def\MeV{{\rm \ MeV}}

\def\Tr{{\rm Tr}\;}

\newcommand{\beq}{\begin{eqnarray}}
\newcommand{\eeq}{\end{eqnarray}}
\newcommand{\beqno}{\begin{eqnarray*}}
\newcommand{\eeqno}{\end{eqnarray*}}
\long\def\beginomit#1\endomit{}
\def\ben{\begin{enumerate}}
\def\een{\end{enumerate}}
\def\bi{\begin{itemize}}
\def\ei{\end{itemize}}

\def\O{{\cal O}}

\def\Tr{{\mbox{Tr}}}
\def\pl {Phys. Lett.}
\def\pr {Phys. Rev.}
\def\np {Nucl. Phys. }
\def\bs{\indent\indent}
\def\O{{\cal O}}

\begin{document}

\begin{titlepage}\begin{center}
\hfill{ SNUTP-95-060}

\hfill{hep-ph/9505283}

\hfill{June 1995}

\vskip 0.7in
{\LARGE\bf\boldmath  The Role of $\Lambda(1405)$ in
Kaon-Proton Interactions}
\vskip 0.4in
{\large Chang-Hwan Lee$^{a,b}$, Dong-Pil Min$^{a,b}$ and Mannque Rho$^{a,c}$}\\
\vskip 0.1in
{\large a) \it Institute for Nuclear Theory, University of Washington}\\
{\large\it Seattle, WA 98195, U.S.A.}\\
{\large b) \it Center for Theoretical Physics and Department of Physics}\\
{\large \it Seoul National University, Seoul 151-742, Korea}\\
{\large c) \it Service de Physique Th\'{e}orique, CEA  Saclay}\\
{\large\it 91191 Gif-sur-Yvette Cedex, France}\\

\vskip 0.6in
{\bf ABSTRACT}\\ \vskip 0.1in
\begin{quotation}
\noindent
S-wave $K^-p$ scattering into various channels near threshold
are analyzed in heavy-baryon chiral perturbation theory with
$\Lambda (1405)$ introduced as an independent field. This is the
approach that predicted the critical density $2\lsim\rho_c/\rho_0\
\lsim 3$ for negatively
charged kaon condensation. We show that chiral perturbation expansion
treating the $\Lambda (1405)$ as elementary is consistent with
{\it all} threshold data including a double-charge-exchange process
suppressed at leading order of chiral expansion in the absence
of the $\Lambda (1405)$. We also discuss S-wave $K^+ p$
scattering phase shifts at low energy.
\end{quotation}
\end{center}\end{titlepage}


\section{Introduction}
\bs
It has been shown \cite{LJMR,LBMR} that chiral perturbation theory
to next-to-next-to-leading order ($\O(Q^3)$) with counterterms fixed
by $KN$ scattering lengths \cite{BS} and kaonic atom data\cite{Gal}
predicts that kaon condensation takes place in dense, compact-star matter
at a density $2\lsim \rho_c/\rho_0\lsim 4$. It was found there that
while the $\Lambda (1405)$ treated as an ``elementary" field \footnote{
Whether or not the notion of $\Lambda (1405)$ as an elementary field
can be justified in chiral perturbation theory for $KN$ scattering is certainly
an open question in view of the proximity of the $\Lambda (1405)$ pole
to the $KN$ threshold \cite{savage}. What we did in \cite{LJMR,LBMR} and what
we shall do in this paper is
to assume that the $\Lambda (1405)$ is elementary and to see how
far one can go in explaining various aspects of low-energy
kaon-nucleon and kaon-nuclear interactions.} plays
a crucial role in $K^- p$ scattering and also in kaonic atoms,
it has very little influence on kaon condensation.
This is mainly because kaon condensation occurs in initially dense neutron
matter and furthermore at a kinematical regime which is far from
the pole position of the $\Lambda (1405)$. In fact, the prediction
of the kaon condensation threshold was found to be surprisingly robust
not only against the influence of the $\Lambda (1405)$ but also against
uncertainties in the parameters of the theory.

An interesting -- and highly pertinent -- question was raised by Weise
in his recent seminar \cite{Weise}: Is the description treating
the $\Lambda (1405)$ as ``elementary" consistent with all low-energy
$KN$ data (e.g. the double-charge-exchange process
$K^-p\rightarrow \pi^+\Sigma^-$)? Weise has proposed a dynamical
coupled-channel treatment of the $\Lambda (1405)$ with a potential
of range $\sim (400 {\mbox{MeV}})^{-1}$ based on a chiral Lagrangian
at $\O(Q^2)$. This dynamical model is found to successfully describe
all of the available low-energy $KN$ data, in particular, the observation that
the OZI-suppressed process $K^-p \rightarrow \pi^+\Sigma^-$ has a lot bigger
cross section than the process $K^-p \rightarrow \pi^-\Sigma^+$
allowed at the tree level.

An immediate question is: Would this dynamical picture
not give a different prediction for kaon condensation?

We shall show in this letter that once the $\Lambda (1405)$ is implemented
as an ``elementary" field with the parameters of the chiral Lagrangian
determined by the S-wave scattering lengths,
the leading order ($\O(Q)$) chiral Lagrangian
can explain most of the branching ratios and low-energy phase shifts
within our framework.
This is not so surprising if one recalls that in the
Skyrme model, the $\Lambda (1405)$ is a bound state
of a kaon and a soliton so that it cannot be described by a sum of
finite series in chiral perturbation theory. A natural approach in the
context of chiral perturbation theory to such a bound
state is to solve a Lippman-Schwinger equation with a potential generated
by chiral perturbation theory using ``irreducible" graphs in the sense
of Weinberg or if one is sufficiently far from the singularity of the bound
state as in the case of kaon condensation,
to introduce it as an elementary field as we do here.
The former corresponds to
Weise's approach and the latter to ours. Our assertion in \cite{LBMR}
was that we are using
chiral Lagrangians in a region far away from the pole position of the
bound state, our approach should be fully justified.

\section{$K^-p$ Scattering}
\bs
We start by writing down the effective chiral Lagrangian that we shall
use in the calculation.
Let the characteristic energy/momentum scale that we are interested in
be denoted $Q$. The standard chiral counting orders the physical amplitudes
as a power series in $Q$, say, $Q^\nu$, with $\nu$ an integer. To leading
order, the kaon-nucleon amplitude goes as ${\cal O}(Q^1)$, to next order
as ${\cal O}(Q^2)$ involving no loops.
Following Jenkins and Manohar\cite{HBF}, we denote the velocity-dependent
octet baryon fields $B_v$, the octet meson fields
$\exp(i\pi_a T_a/f)\equiv\xi$, the velocity four-vector
$v_\mu$ and the spin operator $S_v^\mu$($v\cdot S_v=0,
S_v^2=-3/4$), the vector current $V_\mu=[\xi^\dagger,\partial_\mu\xi]/2$
and the axial-vector current $A_\mu=i\{\xi^\dagger,\partial_\mu\xi\}/2$,
and write the Lagrangian density to order $Q^2$, relevant for the low-energy
scattering, as
\be
{\cal L} &=& \Tr \bar B_v (iv\cdot D) B_v
   + 2 D\Tr \bar B_v S^\mu_v \{A_\mu,B_v\}
   + 2 F\Tr \bar B_v S^\mu_v [A_\mu,B_v]
\nonumber\\
&&      + (\sqrt 2 g_{\Lambda^\star} \bar\Lambda^\star_v \Tr (v\cdot A B_v)
       +h.c. )
\nonumber\\
&& + a_1 \Tr \bar B_v {\cal M}_+ B_v
   + a_2 \Tr \bar B_v B_v {\cal M}_+
   + a_3 \Tr \bar B_v B_v \Tr {\cal M}_+
\nonumber\\
&&+ d_1 \Tr\bar B_v A^2 B_v + d_2 \Tr\bar B_v (v\cdot A)^2 B_v
  +d_3 \Tr\bar B_v B_v A^2 + d_4 \Tr\bar B_v B_v (v\cdot A)^2
\nonumber\\
&& +d_5 \Tr\bar B_v B_v \Tr A^2 + d_6 \Tr\bar B_v B_v \Tr (v\cdot A)^2
\nonumber\\
&&   +d_7 \Tr\bar B_v A_\mu \Tr B_v A^\mu
   +d_8 \Tr\bar B_v v\cdot A \Tr B_v v\cdot A
\nonumber\\
&& +d_9 \Tr\bar B_v A_\mu  B_v A^\mu
   +d_{10} \Tr\bar B_v v\cdot A  B_v v\cdot A\label{lag}
\ee
where D=0.81 and F=0.44, and the covariant
derivative ${\cal D}_\mu$ for baryon fields is defined by
\be
{\cal D}_\mu B_v =\partial B_v +[V_\mu, B_v].
\ee
and ${\cal M}_+ \equiv \xi {\cal M}\xi+ \xi^\dagger {\cal M}\xi^\dagger$
with ${\cal M}=diag(m_u,m_d,m_s)$.

In (\ref{lag}), the $\Lambda (1405)$ -- denoted $\Lambda^\star$ -- is
introduced
as a matter field on the same footing with the usual octet baryons.
The $\Lambda^\star KN$ coupling constant, $g_{\Lambda^\star}^2=0.15$,
 is fixed to give the empirical decay width
$\Gamma_{\Lambda^\star}= 50\ \MeV$ through  $\Lambda^\star\rightarrow
\pi\Sigma$
with $m_{\Lambda^\star}=1405\ \MeV$ \cite{LJMR}.
The explicit symmetry breaking $a_i$ terms can be determined
by the baryon mass splitting and the $\pi N$ sigma term. Here we use
the results of Kaplan and Nelson\cite{KN},
\be
a_1=-0.28, \;\;\; a_2= 0.56, \;\;\; a_3 = -1.1
\nonumber\\
m_u = 6 \MeV,\;\;\; m_d=12 \MeV, \;\;\; m_s=240 \MeV.
\ee
These parameters give  the sigma term
\be
\Sigma_{KN}= -\frac 12 (m_u+m_s) (a_1+2 a_2 + 4 a_3) \simeq 438 \MeV.
\ee
Given the large error bar ($\pm 0.3$) in $a_3$,
this $\Sigma_{KN}$ is consistent with the recent lattice calculations of
Dong \& Liu\cite{DL} (see also Fukugita et al. \cite{fukugita}),
\be
\Sigma_{KN}=\frac{(m_u+m_s)\langle N| (\bar u u+\bar s s)|N\rangle }
                 {(m_u+m_d)\langle N| (\bar u u+\bar d d)|N\rangle }
  \Sigma_{\pi N}
\simeq 450 \pm 30 \MeV.\label{sigmaterm}
\ee
For S-wave scattering, there is no distinction
between $A^2$ and $(v\cdot A)^2$ in (\ref{lag}). So the combination
\be
\bar d_i = d_i + d_{i+1}\;\;\;\; (i: {\rm odd \;\; number})
\ee
enters in the scattering amplitudes.
This means that there are five independent parameters at ${\cal O}(Q^2)$
for S-wave $KN$ scattering.  These parameters need to be fixed for
the equation of state of nuclear star with kaon condensation \cite{LBMR2}.
The transition matrix elements for various channels are summarized
in Appendix.

Now requiring consistency with the experimental $KN$ scattering
lengths\cite{BS}\footnote{We are not concerned with fine-tuning of the
parameters and hence no errors will be quoted.},
\be
a_0^{K^+p} = -0.31 fm, \;\;\; a_0^{K^-p} =-0.67 +i 0.63 fm
\nonumber\\
a_0^{K^+n}=-0.20 fm, \;\;\; a_0^{K^-n}=+0.37 + i0.57 fm,
\ee
we get two constraints on $\bar d_i$ at ${\cal O}(Q^2)$\cite{LJMR},
\be
(\bar d_s -\bar d_v)_{emp} &\approx& (0.05 - 0.06) fm \nonumber\\
(\bar d_s +\bar d_v)_{emp} &\approx& 0.13 fm
\ee
with the isoscalar constants $d_s$ and the isovector constants
$d_v$ defined by
\be
\bar d_s &=& -\frac{1}{2 M_K^2}(m_u+m_s)(a_1+2 a_2+4 a_3)
     +\frac{1}{4 } (\bar d_1+\bar 2 d_3+4 \bar d_5 +\bar d_7)
\nonumber\\
\bar d_v &=& -\frac{1}{2  M_K^2}(m_u+m_s)\; a_1
     +\frac{1}{4 } (\bar d_1+\bar d_7).
\ee
We are therefore left with three independent
parameters to be determined from other experimental data.

\section{Threshold Branching Ratios}
\bs
Once the constants are fixed from some experiments, we could
then make predictions for other physical quantities in low-energy
$KN$ scattering. In particular, the following threshold branching ratios
are of particular interest \cite{Weise}:
\be
\gamma &=& \frac{|{\cal T}_{\pi^+\Sigma^-}|^2}
                {|{\cal T}_{\pi^-\Sigma^+}|^2} \nonumber\\
R_c &=& \frac{\sum_{i=\pi^\pm\Sigma^\mp} |{\cal T}_i|^2 }
             {\sum_{j=\pi^0\Lambda,\pi^0\Sigma^0,\pi^\pm\Sigma^\mp}
             |{\cal T}_j|^2 }
\nonumber\\
R_n &=& \frac{|{\cal T}_{\pi^0\Lambda}|^2}
   {\sum_{i=\pi^0\Lambda,\pi^0\Sigma^0}|{\cal T}_i|^2}
\ee
The empirical values are \cite{Branching}
\be
\gamma^{exp}=2.36\pm 0.04,\;\;\; R_c^{exp}=0.664\pm 0.011,
     \;\;\; R_n^{exp}=0.19\pm 0.02.
\ee
Since we still have three constants left unfixed, we cannot compare theory
directly with experiments for these quantities. However
for the chiral perturbation approach to be viable, the leading
${\cal O}(Q^1)$ chiral
order which involves no unknown counterterms should dominate.
In other words, higher-order terms
should be suppressed according to the counting rule, $(Q/4\pi f)^\nu$.
To verify this, we calculate the branching ratios with the scattering
amplitudes computed at ${\cal O}(Q^1)$. The results are
summarized in Table~\ref{leta}.
\begin{table}
$$
\begin{array}{|c|c|c|}
\hline
 {\rm channel} & f^2 {\cal T}_{\nu=1} & f^2 {\cal T}_{\Lambda^\star} \\
\hline
 \pi^-\Sigma^+  & ( 2 M_K +m_B -m_{\Sigma^+})/4 &  g(m_{\Sigma^+}) \\
 \pi^+\Sigma^-  &    -                           &  g(m_{\Sigma^-}) \\
 K^- p          & M_K                           &  g(m_{p}) \\
 \bar{K^0} n    & M_K /2                        &  g(m_{n}) \\
 \pi^0\Sigma^0  & ( 2 M_K +m_B -m_{\Sigma^0})/8 &  g(m_{\Sigma^0}) \\
 \pi^0\Lambda   & \sqrt{3}( 2 M_K +m_B -m_{\Lambda})/8 &  - \\
 \eta\Sigma^0  & \sqrt{3}( 2 M_K +m_B -m_{\Sigma^0})/8 &  - \\
 \eta\Lambda   & {3}( 2 M_K +m_B -m_{\Lambda})/8 &  g(m_{\Lambda}) \\
 K^+\Xi^- &          -                        &  g(m_{\Xi^-}) \\
 K^0\Xi^0 &          -                        &  g(m_{\Xi^0}) \\
\hline
\end{array}
$$
$$g(m)=-g_{\Lambda^\star}^2\frac{M_K (M_K +m_B-m)}
{m_B +M_K-m_{\Lambda^\star}}$$
\caption{Leading-order contributions to $K^-p$ scattering}
\label{leta}
\end{table}
The numerical results are
\be
\gamma=1.93,\;\;\; R_c=0.64,\;\;\; R_n=0.11.
\ee
Here only the Weinberg-Tomozawa term and the $\Lambda^\star$ contribution
in leading order are taken into account. Clearly the leading tree
contributions play a dominant role for $\gamma$ and $R_c$ while $R_n$
apparently
requires some higher order corrections. Note that without the $\Lambda^\star$,
the transition to $\pi^+\Sigma^-$ would be  suppressed at the leading order, so
we would have $\gamma=0$. Furthermore the enhancement of the $\pi^+\Sigma^-$
channel over the $\pi^-\Sigma^+$ channel is principally due to
$\Lambda^\star$: while only the $\Lambda^\star$ term  contributes to
the $\pi^+\Sigma^-$ channel, both the
Weinberg-Tomozawa term and the $\Lambda^\star$ term contribute
to the channel $\pi^-\Sigma^+$ but with an opposite sign, giving rise to
the  enhancement of the ratio $\gamma$.

{}From the point of view of chiral perturbation theory for kaon condensation,
the most meaningful outcome of the present exercise is that we are now able to
determine the three parameter combinations that appear at $\O (Q^2)$.
This determination would allow us to calculate to $\O (Q^2)$
the equation of state
needed for describing the properties of the kaon condensed state.
The best-fit parameters  using $(\bar d_s -\bar d_v)_{emp}=0.055 fm$
are given in Table~\ref{brat}.

\begin{table}
$$
\begin{array}{|c|c|c||c|c|c|}
\hline
 \bar d_1 + \bar d_7 &\bar d_3 + \bar d_7
  & \bar d_7+ \bar d_9 & \gamma & R_c & R_n \\
\hline
0.039 & -3.35 & -5.90 & 2.36 & 0.63 & 0.19 \\
\hline
\end{array}
$$
\caption{Best-fit parameters iin $fm$ for the branching ratios}\label{brat}
\end{table}

\section{S-Wave Phase Shift for $K^+p$ Scattering}

\begin{figure}
\hskip 1cm
{\epsfig{file=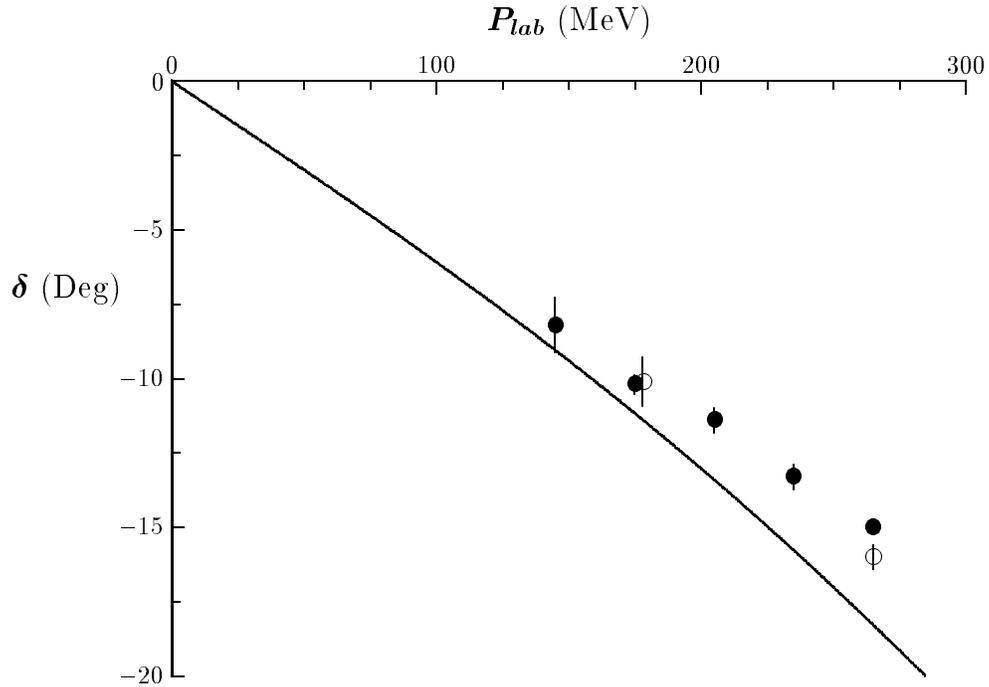,height=9cm}}
\caption{The S-wave phase shifts (in deg) for $K^+p$ scattering.
The filled and empty circles correspond to the data of
\protect\cite{Cameron} and \protect\cite{Burnstein}, respectively. }
\label{fig-phase}
\end{figure}

\bs
While the $\Lambda (1405)$ contributes unimportantly to
the S-wave phase shifts for $K^+p$ scattering
(since it enters in the crossed term
with a large energy denominator), it is however important to check
whether or not the chiral perturbation approach to
kaon-nuclear interactions with the large $KN$ sigma term,
$\Sigma_{KN}\approx 438$ MeV, is consistent with the well-measured
$K^+ p$ phase shifts \cite{Cameron,Burnstein}. We have computed the
phase shifts to ${\cal O} (Q^2)$ chiral order for which there
are no unknown constants once the scattering lengths are fit. The results
are given in Table \ref{tab-phase} and and Fig.~\ref{fig-phase}. Although
there are no experimental data to compare with for $P_{lab} < 145$ MeV,
we see that the ${\cal O} (Q^2)$ chiral perturbation theory is consistent
with the low-energy part of the phase shifts up to $P_{lab}\sim m_\pi$.
The deviation
seen at $P_{lab}\gsim 145$ MeV must clearly be due to higher chiral order
terms.
This will be checked in a future publication at ${\cal O} (Q^3)$ and beyond
using the result of Ref.\cite{LBMR}.

\begin{table}
$$
\begin{array}{|c|c|c|c|c|}
\hline
P_{lab} & \omega_{K,cm} & W.\ Cameron\cite{Cameron}
 & R.A. \ Burnstein\cite{Burnstein} & Our \ Result \\
\hline
145  & 504   & -8.2\pm 0.9   & -             & -9.05 \\
175  & 508   & -10.2\pm 0.3  &  -            & -11.2 \\
178  & 509   &  -            & -10.1 \pm 0.8 & -11.4 \\
205  & 513   & -11.4 \pm 0.4 & -             & -13.4 \\
235  & 518   & -13.3 \pm 0.4 & -             & -15.8 \\
265  & 525   & -15.0\pm 0.2  & -16.0\pm 0.4  & -18.3 \\
\hline
\end{array}
$$
\caption{Comparison of the predicted
phase shifts (in deg) with experimental data.
 $P_{lab}$ and $\omega_{K,cm}$ are given in $MeV$.}
\label{tab-phase}
\end{table}

\section{Discussion}
\bs
We have shown that chiral perturbation theory with the $\Lambda (1405)$
introduced as an elementary matter field can satisfactorily describe
low-energy S-wave $K^{\pm} p$ scattering. In particular, the OZI-suppressed
double-charge-exchange process $K^-p\rightarrow \pi^+\Sigma^-$
is found to be enhanced relative
to the OZI-allowed process $K^-p\rightarrow \pi^-\Sigma^+$ when
$\Lambda (1405)$ is introduced. We have also shown that a large $KN$ sigma
term, $\Sigma_{KN}\approx 430$ MeV, is consistent with
the experimental phase shifts for $K^+p$ scattering at low energy,
$P_{lab}\lsim m_\pi$. Our results provide evidence that the $\Lambda (1405)$
as an elementary field is a phenomenologically viable concept, in a way
that resembles the $\Delta$ in the nonstrange sector.

Finally we reiterate the conclusion of \cite{LBMR}
that the $\Lambda (1405)$ interpolating as an elementary matter
field plays a negligible role in kaon condensation with the subtleties
associated with the threshold properties of $KN$ scattering affecting
little the condensation phenomenon.

\subsection*{Acknowledgments}
\bs
We would like to thank G.E. Brown and W. Weise for helpful discussions:
this work was inspired by the seminar of Weise and discussions following it.
We acknowledge the hospitality and support of
the Institute for Nuclear
Theory at the University of Washington where we were participating
in the INT95-1 program on ``Chiral symmetry in hadrons and nuclei"
and the partial support of the Department of
Energy. CHL would also like to thank Prof. R. Ingalls of
Univ. of Washington and B.Y. Park for hospitality and help.
The work of CHL and DPM was supported in part by the
Korea Science and Engineering Foundation through the CTP of SNU and in
part by the Korea Ministry of Education under Grant No. BSRI-95-2418.

\renewcommand{\theequation}{A.\arabic{equation}}
\section*{Appendix : Transition Amplitudes }
The transition matrix elements for various channels of threshold
$K^-p$ scattering are
\be
{\cal T} (K^-p\rightarrow \pi^-\Sigma^+) &=&
     \frac{1}{4 f^2}  (2 M_K +m_p -m_{\Sigma^+} )
     -\frac{g_{\Lambda^\star}^2}{f^2}\frac{M_K (M_K +m_p-m_{\Sigma^+})}
                             {m_p +M_K-m_{\Lambda^\star}}
     \nonumber\\
     && -\frac{1}{2 f^2} a_2 (2 m_u +m_d+m_s)
     \nonumber\\
     && +\frac{1}{2 f^2} (\bar d_3+\bar d_7) M_K (M_K+m_p-m_{\Sigma^+})
\nonumber\\
{\cal T} (K^-p\rightarrow \pi^+\Sigma^-) &=&
     -\frac{g_{\Lambda^\star}^2}{f^2}
     \frac{M_K (M_K +m_p-m_{\Sigma^-})}
                            {m_p +M_K-m_{\Lambda^\star}}
     \nonumber\\
      && +\frac{1}{2 f^2} (\bar d_7+\bar d_9) M_K (M_K+m_p-m_{\Sigma^-})
\nonumber\\
{\cal T} (K^-p\rightarrow K^- p) &=&
 \frac{1}{f^2} M_K
    -\frac{g_{\Lambda^\star}^2}{f^2}\frac{M_K^2 }
                             {m_p +M_K-m_{\Lambda^\star}}
     \nonumber\\
     && -\frac{1}{f^2}(m_u+m_s)(a_1+a_2+2 a_3)
     \nonumber\\
     && +\frac{1}{2 f^2} (\bar d_1+\bar d_3+2 \bar d_5 +\bar d_7)
     M_K^2
\nonumber\\
{\cal T} (K^-p\rightarrow \bar{K^0} n) &=&
 \frac{1}{2 f^2} M_K
    -\frac{g_{\Lambda^\star}^2}{f^2}\frac{M_K (M_K +m_p-m_n)}
                             {m_p +M_K-m_{\Lambda^\star}}
     \nonumber\\
     && -\frac{1}{2 f^2}(m_u+m_d+2 m_s) a_1
     \nonumber\\
     && +\frac{1}{2 f^2} (\bar d_1+\bar d_7) M_K (M_K+m_p-m_n)
\nonumber\\
{\cal T} (K^-p\rightarrow \pi^0\Sigma^0) &=&
 \frac{1}{8 f^2} (2 M_K +m_p -m_{\Sigma^0})
    -\frac{g_{\Lambda^\star}^2}{f^2}\frac{M_K (M_K +m_p-m_{\Sigma^0})}
                             {m_p +M_K-m_{\Lambda^\star}}
     \nonumber\\
     && -\frac{1}{4 f^2} (3 m_u+m_s) a_2
     \nonumber\\
     && +\frac{1}{4 f^2} (\bar d_3+2 \bar d_7+\bar d_9)
         M_K (M_K+m_p-m_{\Sigma_0})
\nonumber\\
{\cal T} (K^-p\rightarrow \pi^0\Lambda) &=&
     \frac{\sqrt{3}}{8 f^2} (2 M_K +m_p -m_{\Lambda})
     \nonumber\\
     && -\frac{1}{4\sqrt 3 f^2}(3 m_u+m_s)(-2 a_1+a_2 )
     \nonumber\\
     && +\frac{1}{4\sqrt 3 f^2} (-2 \bar d_1+ \bar d_3+\bar d_9)
     M_K (M_K+m_p-m_\Lambda)
\nonumber\\
{\cal T} (K^-p\rightarrow \eta\Sigma^0) &=&
    \frac{\sqrt 3}{8 f^2}(2 M_K +m_p -m_{\Sigma^0})
    \nonumber\\
    && -\frac{1}{4\sqrt 3 f^2} a_2 (m_u-5 m_s)
     \nonumber\\
     && +\frac{1}{4\sqrt 3 f^2} (-\bar d_3+\bar d_9)
     M_K (M_K+m_p-m_{\Sigma^0})
\nonumber\\
{\cal T} (K^-p\rightarrow \eta\Lambda) &=&
    \frac{3}{8 f^2}  (2 M_K +m_p -m_{\Lambda})
    -\frac{g_{\Lambda^\star}^2}{f^2}\frac{M_K (M_K +m_p-m_{\Lambda})}
                             {m_p +M_K-m_{\Lambda^\star}}
    \nonumber\\
    && -\frac{1}{12 f^2} (2 a_1-a_2) (5 m_s- m_u)
     \nonumber\\
     && +\frac{1}{12 f^2} (2\bar d_1-\bar d_3+6 \bar d_7+5 \bar d_9)
     M_K (M_K+m_p-m_{\Lambda})
\nonumber\\
{\cal T} (K^-p\rightarrow K^+\Xi^-) &=&
   -\frac{g_{\Lambda^\star}^2}{f^2}\frac{M_K (M_K +m_p-m_{\Xi^-})}
                            {m_p +M_K-m_{\Lambda^\star}}
     \nonumber\\
      && +\frac{1}{2 f^2} (\bar d_7+\bar d_9) M_K (M_K+m_p-m_{\Xi^-})
\nonumber\\
{\cal T} (K^-p\rightarrow K^0\Xi^0) &=&
   -\frac{g_{\Lambda^\star}^2}{f^2}\frac{M_K (M_K +m_p-m_{\Xi^0})}
                            {m_p +M_K-m_{\Lambda^\star}}
     \nonumber\\
      && +\frac{1}{2 f^2} (\bar d_7+\bar d_9) M_K (M_K+m_p-m_{\Xi^0})
\ee
where the physical masses of baryon and meson are used.
The processes involving $\bar K$, $\eta$ and $\Xi$ do not figure
in the branching ratios $R_c$ and $R_n$\cite{Branching}.


\end{document}